\documentclass[12pt]{article}

\usepackage{fullpage,times,amsmath,graphicx,amsthm,amssymb,amsfonts,enumerate,multirow,setspace,bm,bbm,hyperref}

\usepackage{natbib}
\usepackage{rotating}
\usepackage[labelfont=bf]{caption}

\usepackage{algorithm2e}
\usepackage{algorithmic}
\usepackage{graphics,epsfig,color}

\usepackage{booktabs}

\newcommand{\E}{\ensuremath\mathrm{E}}

\title{Better lower bounds for missing species:
improved non-parametric moment-based estimation for large experiments}

\author{Timothy Daley \\ email: \href{mailto:tdaley@stanford.edu}{tdaley@stanford.edu}
\and Andrew D Smith \\ email: \href{mailto:andrewds@usc.edu}{andrewds@usc.edu}}

\date{}

\begin{document}

\maketitle

\begin{abstract}
Estimation of the number of species or unobserved classes from a
random sample of the underlying population is a ubiquitous problem in
statistics.  In classical settings, the size of the sample is usually
small.  New technologies
such as high-throughput DNA sequencing have allowed for the sampling of
extremely large and heterogeneous populations at scales not previously
attainable or even considered.  New algorithms are required that take
advantage of the size of the data to account for heterogeneity, but
are also sufficiently fast and scale well with large data.
We present a non-parametric moment-based estimator that is both
computationally efficient and is sufficiently flexible to account for
heterogeneity in the abundances of underlying population.
This estimator is based on an extension of a popular moment-based lower
bound~\citep{chao1984nonparametric}, originally developed by
\cite{harris1959determining} but unattainable due to the lack of
economical algorithms to solve the system of nonlinear equation
required for estimation.
We apply results from the classical moment problem to show that
solutions can be obtained efficiently, allowing for estimators that
are simultaneously conservative and use more information. 
 This is critical for modern genomic applications, where
there may be many large experiments that require 
the application of species estimation.
 We present applications of our
estimator to estimating 
T-Cell receptor repertoire and 
dropout in single cell RNA-seq experiments.
\end{abstract}

\section{Introduction}
\label{sec:intro}

Consider the problem of the estimating the number of distinct classes
in an unknown population through sampling. Researchers sample
instances from a finite and closed population and use the pattern of
repeated sampling to estimate the number of unsampled classes. This
problem is ubiquitous, with applications in
immunology~\citep{qi2014diversity},
microbiology~\citep{bunge2014estimating},
epidemiology~\citep{bohning2005nonparametric},
linguistics~\citep{efron1976estimating},
genetics~\citep{gravel2014predicting}, database
management~\citep{haas1995sampling}, among others.  See
\cite{deng2019molecular} for a review of applications arising from
high-throughput DNA sequencing data.

The original, and still primary, use of this problem was in
ecology~\citep{fisher1943relation} and most of the language and
terminology arise from this application.  The distinct classes are
known as species and the instances are known as individuals.  In such
experiments, captures may be
are costly and difficult.  The total number of samples is typically in
the hundreds or thousands.  For example in \cite{fisher1943relation},
620 species of butterfly were captured from the Malay peninsula,
and more recently \cite{gopalakrishnan2018gut} captured 1,436
microbial species (as determined by operational taxonomic units) in human
patients.
The small sample size results in very few degrees of freedom for
identifying and estimating heterogeneity in the population.
Estimators tend to focus on mild heterogeneity, such as simple
contamination that results in small deviations from uniform
sampling~\citep{bohning2013use,chiu2014improved}. 
When the number of samples is small, then approximating the
non-uniformity with simple distributions may be appropriate because
of the fact that it's difficult to model complex heterogeneity
when the data is small.
On the other hand, when the sample size is large, say in the millions
or billions, then researchers have a better opportunity to account for
this heterogeneity and better estimate the number of unobserved
species.

In our applications of interest,
which arise from high-throughput sequencing experiments that sample
millions or billions of DNA molecules,
uniformity of the population is a hopeless thought.
There are a myriad of inherent biases from the sequencing technology
that vary between samples
\citep{sims2014sequencing},
in addition to the fact that the most interesting biology to study
are typically the ones that inherently contain a large number of extremely rare classes
\citep{bunge2014estimating,qi2014diversity}.
Consider the following example: T-cell receptor (TCR) $\beta$ CDR3
clonotypes were interrogated in 39 individuals of various ages using
high-thoughput sequencing~\citep{britanova2014age}.  In one sample,
presented in Table~\ref{tcr_freq}, a total of 992,830 TCR $\beta$ CDR3
sequences were sampled with 705,621 distinct clones (or classes).  The
other samples in the study are also of similar size.  The need for
fast and accurate estimation is required for further analysis, such as
age and environmental related patterns in TCR diversity.

\begin{table}[t!]
\begin{center}
\begin{tabular}{|r|r|r|r|r|r|r|r|r|r|r|r|}
\hline
$j$ & 1 & 2 & 3 & 4 & 5 & 6 & 7 & \ldots & 3400 & 7288 & 7733  \\
\hline
$n_{j}$ & 603,776 & 73,628 & 14,113
& 3,691 & 2,446 & 1,612 &
1, 148 & \ldots & 1 & 1 & 1 \\
\hline
\end{tabular}
\caption{Example T-cell  $\beta$ receptor
count frequencies in a healthy
11 year old female.}
\label{tcr_freq}
\end{center}
\end{table}

There are several estimators currently available
available for researchers.
The non-parametric maximum likelihood estimator (NPMLE)
of \cite{norris1998non} 
fits the discrete MLE under a Poisson model.
Due to the inherent instability of this procedure that may result
in a component with arbitrarily small abundance,
\cite{wang2005penalized} 
developed a penalized version of the non-parametric
MLE, as the penalization
helps to prevent the boundary problem in the NPMLE.
\cite{willis2015estimating} 
presented a method based on non-linear
regression of successive frequency ratios.
\cite{efron1976estimating} 
derived a linear programming based lower bound.
\cite{burnham1978estimation} developed a jackknife-based
estimator.
But the most widely used estimator
is a moment-based non-parameteric lower
bound from~\cite{chao1984nonparametric}.
This estimator is easy to
compute and is fairly robust to small deviations from
homogeneity~\citep{chao1977quadrature}.  Unfortunately, it only uses a
small amount of information of the sample, just the number of species
observed exactly once and twice, and will severely underestimate the
population size when the population is highly heterogeneous.

Here we present a moment-based non-parametric method to extend Chao's
lower bound by taking into account more information available in the
experiment.  We first note that in a Poisson or multinomial mixture
model, Chao's lower bound is a special case of a general framework
proposed by \cite{harris1959determining}.  In this framework the
problem of estimating the number of unobserved classes is reformulated
to the problem of estimating an integral of an unknown measure and the
information in the sampled counts is reformulated into information on
the unknown measure's moments.

The problem of estimating an integral over an unknown measure with
specified moments is well known in the numerical analysis literature
and has a solution given by Gaussian quadrature.  Gaussian quadrature
can produce upper and lower bounds to the integral of interest using
discrete approximations that are calculated via moment matching and
the technique is intimately related to the classical moment problem
\citep{golub2010matrices}.
This connection between the two problems allows us to extend the
theory of the classical moment problem and advanced algorithms of
Gaussian quadrature specifically for the problem of estimating the
number of unobserved classes in a population.
We will show that this allows us to obtain estimates that improve upon
the Chao estimator in diverse and heterogeneous populations. 

We present two applications of our estimator to 
problems in modern genomics: 
estimation of TCR repertoire, where
species abundance is naturally highly heterogeneous, 
and estimation of
dropout for single cell RNA-seq (scRNA-seq)
experiments.
In the former case we show that patterns in TCR repertoire
may be missed using the observed repertoire
that appear in the species-corrected repertoire.
In the latter case, dropout is a consequence of 
the single cell sampling technology and can 
bias downstream analysis.
We show in simulations 
that taking into account the species-corrected
dropout in the differential expression analysis
can improve identification of truly differentially 
differentially expressed genes.

The remainder of the paper is organized as follows.
Section~\ref{sec:model} introduces
the underlying model and problem.
Section~\ref{sec:mom_spaces}
introduces our moment-based solution to estimating the number of
unobserved classes and discusses Gaussian quadrature algorithms for
computing the estimator.
Section~\ref{sec:eval} shows
results and comparisons to other estimators in the simple case where
the abundance distribution is a discrete mixture distribution.
Section~\ref{sec:bagging} discusses a bagging strategy to alleviate an
ill-conditioning issue with our estimator.
Section~\ref{sec:app} shows
two applications of estimating the number of classes in
high-throughput biology.
We conclude with some remarks and discussions in
section~\ref{sec:discuss}.
Throughout the paper, for the sake of brevity we state the necessary
theory without proofs, referring the reader instead to the requisite
references.

\section{Model}
\label{sec:model}

Consider a population of $S < \infty$ classes or species.  In practice
$S$ is an integer, but in the theory below we will let it be any
number between zero and infinity.  Let $y_{i}$ be the number of
sampled individuals from class $i$, for $i = 1, \ldots, S$, and let $N
= \sum_{i = 1}^{S} y_{i}$ be the total number of sampled individuals,
sometimes called the sampling effort.  We assume that $\boldsymbol{y}
= \{ y_{i}; i = 1, \ldots, S \}$ arise as independent Poisson random
variables with rates $\boldsymbol{\lambda} = \{ \lambda_{i}; i = 1,
\ldots, S \}$.  In the applications we consider, where the samples
originate from high-throughput sequencing technologies, this
assumption is appropriate since the total number of individuals
sampled is inherently random.

If all species are equally abundant, then
$\lambda_{1} = \lambda_{2} = \ldots = \lambda_{S}$.
This situation is well-studied in the literature~\citep{bunge1993estimating},
but the equal abundance assumption is not likely to hold
in practice.
We assume that the Poisson rates are independently and
identically distributed according to some latent distribution
$\mu (\lambda)$,
so that the sampled class counts will follow a Poisson mixture model.
The latent mixing distribution encompasses all factors that may affect the
abundance of classes,
including both fundamental factors, such as the underlying
biology~\citep{desponds2016fluctuating},
and technical factors,
such as PCR amplification bias~\citep{benjamini2012summarizing}.

Define $f_{\mu} (j) = \int_{0}^{\infty} e^{- \lambda} \lambda^{j} / j!
\, d \mu (\lambda)$ to be the probability that a randomly
chosen class is sampled exactly $j$ times.
Let the count frequency $n_{j}$ be the number of
classes with exactly $j$ individuals sampled,
$n_{j} = \sum_{i = 1}^{S} \mathbbm{1} \big(y_{i} = j \big)$
with expectation equal to
\[
\E \big( n_{j} \big) = S f_{\mu} ( j ) =
S \int_{0}^{\infty} e^{ - \lambda} \lambda^{j} / j! \, d \mu (\lambda).
\]
Note that $n_{0}$ is the number of unobserved classes
and since $S = n_{0} + \sum_{j = 1}^{N} n_{j}$,
estimating $n_{0}$ is equivalent to estimating $S$.


\subsection{Harris' change of measure}
\label{sec:estim}

Following an idea presented by \cite{harris1959determining} and
\cite{chao1984nonparametric}, we define the measure $\nu$ as a
transformation of the abundance distribution $\mu$ such that
\[
d \nu (\lambda) = S \, \lambda e^{-\lambda} d \mu (\lambda).
\]
Note that $\nu$ is not a probability measure since it is
unnormalized with
$\int_{0}^{\infty} d \nu(\lambda) = \mathrm{E} n_{1} $.
By definition this is a bijection with
$d \mu (\lambda) = S^{-1} \,  \lambda^{-1} e^{\lambda} d \nu (\lambda)$ and
the support of $\mu$ and $\nu$ are identical.

The main benefit of defining and working with the transformed
measure $\nu$ is that the moments of $\nu$
can be expressed as simple functions of the expected counts
frequencies.
Specifically the expected moments of $\nu$ can be derived as
\begin{equation}
\nu_{m} = \int_{0}^{\infty} \lambda^{m} d \nu (\lambda) = S
  \int_{0}^{\infty} \lambda^{m + 1} e^{ - \lambda} d \mu (\lambda)
= (m + 1)! \, \E \big( n_{m + 1} \big),
\label{nu_moments}
\end{equation}
with $\hat{\nu}_{m} = (m + 1)! n_{m + 1}$ denoting the
estimated moments by plugging in the observed count frequencies
for the expected count frequencies.
Furthermore, we can express $\E (n_{0})$ in terms of the modified
measure $\nu$ as
\begin{align}
\E \big( n_{0} \big) &=
S \int_{0}^{\infty} e^{- \lambda} d \mu (\lambda)
= \int_{0}^{\infty} \frac{1}{\lambda} \, S \lambda e^{-\lambda} d \mu(\lambda)
\notag \\
&=  \int_{0}^{\infty} \lambda^{-1} d \nu (\lambda).
\label{nu_unobs}
\end{align}
This expression is free of $S$, since it has been absorbed
into the unnormalized measure $\nu$.
The estimation of $\E (n_{0} )$ is now equivalent to
estimation of the integral $\int_{0}^{\infty} \lambda^{-1} d \nu (\lambda)$.

The information we have on $\nu$ from the data is through the moments,
which we can estimate by using the observed count frequencies.
We now state the problem of estimating the number of
missing species $n_{0}$ in the form of
a moment-constrained problem as follows:
\begin{align} \label{problem}
\text{Estimate}~&\int_{0}^{\infty} \lambda^{-1} d \nu (\lambda) \\
\text{such that}~&\hat{\nu}_{m} =
(m + 1)! n_{m+1}~\text{for}~m = 0, 1, \ldots, M. \notag
\end{align}
We will refer to $M$ as the order of the estimate.

\section{Moment spaces}
\label{sec:mom_spaces}

Consider the space of all measures on the positive real line that
satisfy the expected moment constraints in equation~\eqref{problem},
\[
\mathcal{V}_{M} = \{ \xi : \xi \text{ is a measure on the
positive reals and }\xi_{m} =
\nu_{m}~\text{for}~ m = 0, 1, \ldots, M \},
\]
called a $M$-truncated moment space.
$\mathcal{V}_{M}$ is closed and convex
~\citep[Theorem 1]{harris1959determining}.
For example, if $\nu_{0} = 1000$ and $\nu_{1} = 1000$  then
$\mathcal{V}_{1} = \{ \xi : \xi_{0} =  1000, \xi_{1} = 1000 \}$ and
encompasses all measures with the first
two moments equal to 1000.
Examples of the pair $(S, \mu)$
that satisfy these two moment conditions and
are contained in $\mathcal{V}_{1}$ are
the pair $S = 4000$ and $\mu$ equal to a Gamma$(k = 1, \theta = 1)$,
and the pair $S = 1000 e^{1}$ and
$\mu$ equal to the single point $\lambda = 1$,
as well as any convex combination of these.

By transforming the measure $\mu$ to the measure $\nu$,
the problem of estimating the number of missing species
is equivalent to understanding
how the functional $\int_{0}^{\infty} \lambda^{-1} d \eta (\lambda)$
acts within the space $\mathcal{V}_{M}$.
Classical resources for this problem include
\cite{karlin1953geometry} and \cite{karlin1966tchebycheff}.
We will summarize key results below.

Let $P$ denote the number of support points
of $\nu$, with $P = \infty$ if $\nu$ has continuous support.
Then for $M > 2P - 1$, $\mathcal{V}_{M}$ contains
only a single point, namely $\nu$.
On the other hand, if $M \leq 2P - 1$ then $\mathcal{V}_{M}$
contains not only $\nu$, but also other measures whose
first $M$ moments correspond with the first $M$ moments of $\nu$.

If $P = \infty$ and $S < \infty$ then the moments of $\nu$
are all bounded, {\em i.e.} for some constant $C$
independent of $m$
\[
\nu_{m} = \int_{0}^{\infty} \lambda^{m} d \nu (\lambda)
= S \int_{0}^{\infty} \lambda^{m + 1} \sum_{k = 0}^{\infty}  \frac{1}{k! \lambda^{k}} d \mu(\lambda)
\leq C \int_{0}^{\infty} d \mu(\lambda) = C.
\]
$\mathcal{V}_{M}$ converges
to a single point as $M \to \infty$
\citep[Proposition 1.5]{simon1998classical}.
This in turns implies the identifiability of the
Poisson mixture model.
In other words, if all of the expected count frequencies
$\{ \E (n_{j}); j = 1, 2, \ldots \}$ are all known to the
researcher, then $\mu$ can be
perfectly recovered.
See \cite{mao2001moment} or \cite{mao2007estimating}
for similar results in the context of the missing species problem.

If only a finite number of the count frequencies are known (as will
always happen in practice due to finite sampling effort) then
$\mathcal{V}_{M}$ contains all measures with support $P \geq (M +
1)/2$ whose first $M$ moments coincide with the first $M$ moments of
$\nu$.
Indeed, we can show that when $M$ is finite
and when $\mathcal{V}_{M}$ contains at least one measure that satisfy the moment
conditions, then there are an infinite number of measures that
do.
This includes two special discrete distributions in particular,
one that gives a lower bound to the integral in
equation~\eqref{problem} and one that gives an
upper bound.
In the former case, the support will be strictly in the interval
$(0, \infty)$, and this results in a lower bound that is finite.
In the latter case, the support will contain the point $0$,
and this results in an infinite upper bound.
This implies that without an infinite number of non-zero
moments (and an infinite amount of data),
we cannot exclude the possibility that there are
an infinite number of species in the population.
This is reminiscent of similar results by
\cite{wang2005penalized} and
\cite{mao2007estimating}.
The former showed that the non-parametric maximum
likelihood estimator can include $0$ in the support and
the latter showed that upper confidence intervals for the missing species
always has a non-zero probability of including zero.
Indeed, this indicates that without severe restrictions on the
abundance distribution we can not exclude the possibility that there
an infinite number of classes in the population.
On the other hand, we can always obtain finite lower bounds,
which may default to the Chao lower bound in the worst case.


\subsection{Moment-based estimation}
\label{sec:mom_est}

When $M$ is less than the two times the number of support points
of $\mu$ minus one
({\em i.e.} $M < 2P - 1$, where $P$ is the number of support points of $\mu$),
the set $\mathcal{V}_{M}$ is closed, convex, and bounded
\citep{karlin1953geometry}.
If $g(x)$ is a strictly convex or
concave function (and $g(x) = x^{-1}$ is the former) then the linear
functional $\mathcal{L}$ defined by
\[
\mathcal{L} (\xi) = \int_{0}^{\infty} g(\lambda) d \xi(\lambda)
\]
will attain its minima and maxima on the boundary of
$\mathcal{V}_{M}$.
The boundary consists of discrete measures
of minimal degree that satisfy the moment constraints~\citep{harris1959determining}.
For the lower bound, this is a measure $\hat{\nu}$
with support
$0 < \psi_{1} < \psi_{2} < \ldots < \psi_{P} < \infty$
and corresponding weights
$\{ \omega_{i}:  0 < \omega_{i} < \infty, i = 1, \ldots, P \}$
that satisfy the following system of equations:
\begin{align}
\omega_{1} + \ldots + \omega_{P} &= \hat{\nu}_{0}
\notag \\
\omega_{1} \psi_{1} + \ldots + \omega_{P} \psi_{P} &= \hat{\nu}_{1}
\notag \\
&\vdots
\notag \\
\omega_{1} \psi_{1}^{M} + \ldots + \omega_{P} \psi_{P}^{2P - 1} &= \hat{\nu}_{2P -1}.
\label{system_eqns}
\end{align}
Assuming we can solve the above system of equations,
an estimated lower bound to the number of unobserved classes is then simply
given by plugging in $\hat{\nu}$ to the integral in
equation~\eqref{problem},
\begin{equation} \label{discrete_estimator}
\hat{n}_{0} = \sum_{i = 1}^{P} \omega_{i} / \psi_{i}.
\end{equation}

\subsection{Existence of solutions}
\label{sec:existence}

For a given $P$,
consider the $P \times P$ moment Hankel matrix defined by
\[
H_{P} = (\nu_{i + j -2} )_{1 \leq i, j \leq P}
= \begin{pmatrix}
\nu_{0} & \nu_{1} & \ldots & \nu_{P - 1} \\
\nu_{1} & \nu_{2} & \ldots & \nu_{P} \\
\vdots & & \ddots & \vdots \\
\nu_{P - 1} & \nu_{P} & \ldots & \nu_{2P - 2}
\end{pmatrix}
\]
and the $P \times P$ shifted moment Hankel matrix defined by
\[
H^{\prime}_{P} = (\nu_{i + j - 1})_{1 \leq i, j \leq P}
= \begin{pmatrix}
\nu_{1} & \nu_{2} & \ldots & \nu_{P} \\
\nu_{2} & \nu_{3} & \ldots & \nu_{P + 1} \\
\vdots & & \ddots & \vdots \\
\nu_{P} & \nu_{P + 1} & \ldots & \nu_{2P - 1}
\end{pmatrix}.
\]
These matrices are critical to the moment problem.
Specifically, the positive definiteness of
$H_{P}$ and $H^{\prime}_{P}$ are both necessary and
sufficient for the existence of a measure $\hat{\nu}$ with moments
$\nu_{0}, \nu_{1}, \ldots, \nu_{2P - 1}$ and support size $= P$
\citep[Theorem 2.1]{pozza2019algebraic}.

To see why the positive definiteness of the Hankel matrix
is necessary, consider the following example.
Define a quadratic form for $\boldsymbol{x} \in
\mathbb{R}^{P}$ by
\begin{align}
\boldsymbol{x}^{T} H_{P} \boldsymbol{x} &= \sum_{i,j = 1}^{P} x_{i} x_{j} \nu_{i + j - 2}
= \sum_{i, j = 1}^{P} x_{i} x_{j} \int_{0}^{\infty} \lambda^{i + j -2} d \nu (\lambda) \notag \\
&=
\int_{0}^{\infty} \Big( \sum_{i = 1}^{P}  x_{i} \lambda^{i} \Big)^2 d \nu(\lambda) > 0,
\notag
\end{align}
where the last equality follows if $\nu$ is a measure over the positive
real numbers.
If the Hankel matrix is not positive definite, then there exists some
$\boldsymbol{x}$ for which the above quadratic form is negative
and we obtain a contradiction.
In a similar manner, the necessity of the positive definiteness of the shifted
moment Hankel matrix can be shown.
Sufficiency, on the other hand, is difficult to show
and we refer readers to \cite{pozza2019algebraic}
for full details and proofs.
Therefore, if the moment and shifted moment
Hankel matrices are not positive definite for some fixed $P$,
then no measures are contained in $\mathcal{V}_{2P - 1}$
and the moment space must be truncated until the Hankel
matrices are positive definite.

Note that in practice we do not observe the true Hankel and shifted
Hankel moment matrices.  We instead observe estimated moment matrices
with the expected moments replaced by their estimates via the count
frequencies, $\hat{\nu}_{m} = ( m + 1)! n_{m + 1}$. Consequently there
is no guarantee that either estimated matrix is positive definite.
This allows us to construct a simple method to choose the order of the
approximation by iteratively checking that the determinants of the
moment Hankel and shifted moment Hankel matrices are greater than some small positive
threshold, with positive definiteness ensured by Sylvester's
criterion.

\subsection{Gaussian quadrature}
\label{sec:gauss_quad}

Now that we have shown the conditions for existence of
solutions exist to the system of equation~\eqref{problem},
we can discuss algorithms for solving it.
We directly apply a modern general non-linear equation solver
such as R package nleqslv
\citep{hasselman2018package} to find a solution,
but this can be problematic.
First, this can be computational intensive
for a large number of moments.
Secondly, and more importantly,
that although
we showed in section~\ref{sec:existence}
that there is only one solution to the
system of equations~\eqref{system_eqns},
this does not prevent the existence of solutions
that may satisfy the system of equations
to numerical precision but are not close to
true solution.
Indeed, one example was shown by \cite{gautschi1983and}.
This is a concern because, as we will discuss later,
the mapping from moments to the quadrature rules
is extremely ill-conditioned, meaning that small changes in
the input can lead to extremely large changes in the
output.
One consequence of this is that the space of numerical
solutions can be large and it's difficult to
check the accuracy of the solutions
\citep{gautschi1983and}.

The moment problem has a rich history and theory associated with it.
We would be remiss to not use this theory to construct efficient algorithms.
Specifically, Gaussian quadrature is a technique that is intimately related
to the moment problem.
It is used to estimate an integral over an
unknown measure given estimates of the moments of the measure
\citep{golub2010matrices}.  Gaussian quadrature constructs a discrete
estimate to the integral such that the estimate is exact for
polynomials up to specified degree.  This ensures that
the discrete approximation satisfies the moment conditions given
in equations~\eqref{problem}.
Suppose that the first $M$
moments of the measure are known.
Let $\psi_{1}, \ldots, \psi_{P}$ be the support of the approximation $\hat{\nu}$
and $\omega_{1}, \ldots, \omega_{P}$ be the weights.
Let $g(\lambda) = \lambda^{m}$ be a monomial.
To construct an exact approximation
then the support and weights must satisfy
\[
\int_{0}^{\infty} \lambda^{m} d \nu(\lambda) =
\omega_{1} \psi_{1}^{m} + \ldots + \omega_{P} \psi_{P}^{m} = \nu_{m}.
\]
To be exact for polynomials up to degree $M$, the above equation
must be satisfied for $m = 0, \ldots, M$,
which is exactly the system of equations~\eqref{system_eqns}
given in section~\ref{sec:mom_est}.

This allows us to apply well known algorithms for
computing Gaussian quadrature rules to solve the system
of equations~\eqref{system_eqns} and construct a moment
based estimator the number of missing species.


The current state of the art algorithms for computing
Gaussian quadrature rules involve estimating
the orthogonal polynomials associated with
the underlying measure.
We have found that this results in more
stable estimates than with non-linear equation solvers
(Figure~\ref{QuadratureOrthPolyVsNleqslv}).

Below we will present the theory specific to our problem
needed to construct efficient algorithms based on
Gaussian quadrature.
For readers interested in a more in-depth explanation,
we suggest the extensive textbooks
of \cite{gautschi2004orthogonal} or
\cite{golub2010matrices}.

Associated with every measure is a unique system of monic polynomials
$\{ p_{m} (\lambda); p_{m} (\lambda) = \lambda^{m} + \ldots, \, m = 0, \ldots \}$
that are mutually orthogonal under the measure $\nu$,
{\em i.e.} $\int_{0}^{\infty} p_{i}(\lambda) p_{m}(\lambda) d \nu (\lambda)$
is zero when $i \neq m$ and strictly positive when $i = m$.
These orthogonal polynomials satisfy a three term recurrence
\begin{align}
&p_{m + 1} (\lambda) = (\lambda - \alpha_{m}) p_{m} (\lambda) - \beta_{m} p_{m - 1} (\lambda),
\label{three_term} \\
&\text{for } \alpha_{m}, \beta_{m} > 0
\text{ and } p_{-1} (\lambda) = 0, \, p_{0} (\lambda) = 1.
\notag
\end{align}
The coefficients $\{\alpha_{m}, \beta_{m} \}_{m \geq 0}$
define the orthogonal polynomials and are important
because of a relationship between the coefficients of the three term
recurrence and the points and weights that solve the system of
equations~\eqref{system_eqns}.

Given the three term recurrence coefficients,
define the Jacobi matrix to be the tridiagonal matrix with
$\alpha_{m}$ on the diagonals and $\sqrt{\beta_{m}}$ on the
off-diagonals.
The truncated Jacobi matrix $J_{P}$ is the $P \times P$ matrix
\[
\begin{pmatrix}
\alpha_{0} & \sqrt{\beta_{1}} & 0 & \ldots & 0 \\
\sqrt{\beta_{1}} & \alpha_{1} & \sqrt{\beta_{2}} & \ldots & 0 \\
\vdots & & & \ddots & \vdots \\
0 & & \ddots & \alpha_{P - 2} & \sqrt{\beta_{P-1}}  \\
0 & & \ldots & \sqrt{\beta_{P-1}} & \alpha_{P-1}
\end{pmatrix}.
\]
The quadrature points $\psi_{1}, \ldots, \psi_{P}$ are the eigenvalues
of $J_{P}$ and quadrature weights are the square of the first
component of the eigenvectors~\citep{golub1969calculation}.
Given the estimated three term recurrence coefficients,
these can be efficiently calculated with a modified QR algorithm
that has complexity that is linear in both space and time.
This is because since $J_{P}$ is already tridiagonal,
the usual Householder transformations
are not needed and we need only the first entry of the eigenvectors.
Thus, it's not necessary to store the full eigenvectors.
Furthermore, this algorithm is perfectly conditioned, meaning
that small changes in the estimated three term recurrence
coefficients will not lead to large changes in the estimated
quadrature points and weights.

\begin{algorithm}[t!]
\caption{The unmodified Chebyshev algorithm}
\begin{algorithmic}[1]
\REQUIRE Estimated moments, $\{ \hat{\nu}_{m}, m = 0, 1, \ldots, 2P - 1\}$,
of an unknown measure $\nu$.
\ENSURE Estimated three term recurrence $\alpha_{0}, \ldots, \alpha_{P - 1}$
and $\beta_{1}, \ldots, \beta_{P - 1}$.
\STATE $\alpha_{0} \leftarrow 1 + \hat{\nu}_{1} / \hat{\nu}_{0}$,
\STATE $\beta_{0} \leftarrow \hat{\nu}_{0}$
\FOR {$l = 1, 2, \ldots, 2P -2$}
\STATE $\sigma_{-1, l} \leftarrow 0$
\FOR {$l = 0, 1, \ldots, 2P - 1$}
\STATE $\sigma_{0, l} \leftarrow \hat{\nu}_{l}$
\ENDFOR
\ENDFOR
\FOR{$k = 1, \ldots, P - 1$}
\FOR {$l = k, k + 1, \ldots, 2P - k - 1$}
\STATE $\sigma_{k, l} \leftarrow \sigma_{k - 1, l + 1} -
\alpha_{k -1} \sigma_{k - 1, l} - \beta_{k - 1} \sigma_{k - 2, l}$
\ENDFOR
\STATE $\alpha_{k} \leftarrow \frac{\sigma_{k, k + 1}}{\sigma_{k, k}}
- \frac{\sigma_{k - 1, k}}{\sigma_{k - 1, l - 1}}$
\STATE $\beta_{k} \leftarrow \frac{\sigma_{k, k}}{\sigma_{k - 1, k - 1}}$
\ENDFOR
\end{algorithmic}
\label{unmod_cheb_algorithm}
\end{algorithm}

On the other hand, the three term recurrence coefficients
can be written as functions of determinants
of matrices that are structurally similar to the moment
Hankel matrices (full details can be found
in \cite[Chapter 2]{gautschi2004orthogonal}).
These matrices have absolute condition number that
grow exponentially in $P$
\citep{gautschi1982generating,tyrtyshnikov1994bad}, and therefore
estimating the three term recurrence coefficients is
extremely ill-conditioned.
In our problem we do not have the expected moments,
instead only noisy estimates of the moments.
A consequence of the ill-conditioning issue is that
due to errors in estimating the moments,
we may have difficulties in accurately estimating the quadrature points.
This is particularly difficult for the smallest point,
which has largest effect on
the estimated number of missing species.

The standard algorithm for calculating the three term recurrence of
the orthogonal polynomials using only moment information is the
modified Chebyshev algorithm~\citep{gautschi2004orthogonal},
originally developed by \cite{sack1971algorithm}.  The modified
Chebyshev algorithm uses a known measure with an analytically derived
three term recurrence to help condition the unknown measure and
requires $O(P^{2})$ computations, rather than $O(P^{3})$ required from
direct computation from the determinant relations.  As we mentioned
previously, the map from the moments to the recurrence coefficients of
the orthogonal polynomials is ill-conditioned and the hope in using
the modified Chebyshev is that a good choice of the modifying measure
will improve the conditioning of the algorithm and improve
estimation~\citep{gautschi1985orthogonal}.  For example, if the known
measure is serendipitously chosen to be unknown measure then the
measure is perfectly recovered.  We previously~\citep{daley2014non}
investigated a case where the true measure is known and the recurrence
coefficients can be calculated analytically, specifically when $\mu$
is a Gamma distribution and the counts are Negative Binomial.  We
found that in the presence of error, the modified Chebyshev algorithm
does not improve performance.  We found instead that the use of no
conditioning measure performs the best in our problem.  Using no
modifying measure results in what is known as the unmodified Chebyshev
and is described in algorithm~\ref{unmod_cheb_algorithm}.  This uses
only the observed moments as input and requires no other input from
the user to calculate the estimated recurrence relation, and to
subsequently calculate the quadrature rules.

\section{Evaluation on discrete distributions}
\label{sec:eval}

We evaluated our moment-based estimator first on
the simplest case of heterogeneity in the abundance distribution,
discrete mixtures.
For this initial comparison,
we tested our method (preseq) against the Chao
estimator (chao; Chao, 1984),
the non-parametric maximum likelihood estimator
(npmle; Norris and Pollock, 1998),
and the penalized non-parametric maximum likelihood
estimator (pnpmle; Wang and Lindsay, 2005).
We use the implementations of the above algorithms
available in the R package SPECIES~\citep{wang2011species},
with the bootstrap option turned off.
For all of the below simulations, the total number of
species is $S = $1,000,000.

We measured the performance of the estimators
in multiple ways.
It is important that the estimator is close on average,
which we measure by the median.
The median ignores the variability of the estimator,
so we used the root mean square error (RMSE)
to measure the bias-variance tradeoff of the estimators.
Since our simulated populations are orders of magnitude larger
than populations previously investigated, we also
measured the running time of the algorithms.
Finally, for the preseq, npmle, and pnpmle methods,
we kept track of the number of times the
methods selected the correct number of components
in the mixture to measure the ability of the algorithms to
choose the correct model.

\subsection{Two component mixture abundance distributions}
\label{sec:2comp}

\begin{figure}[t!]
\centering{\includegraphics{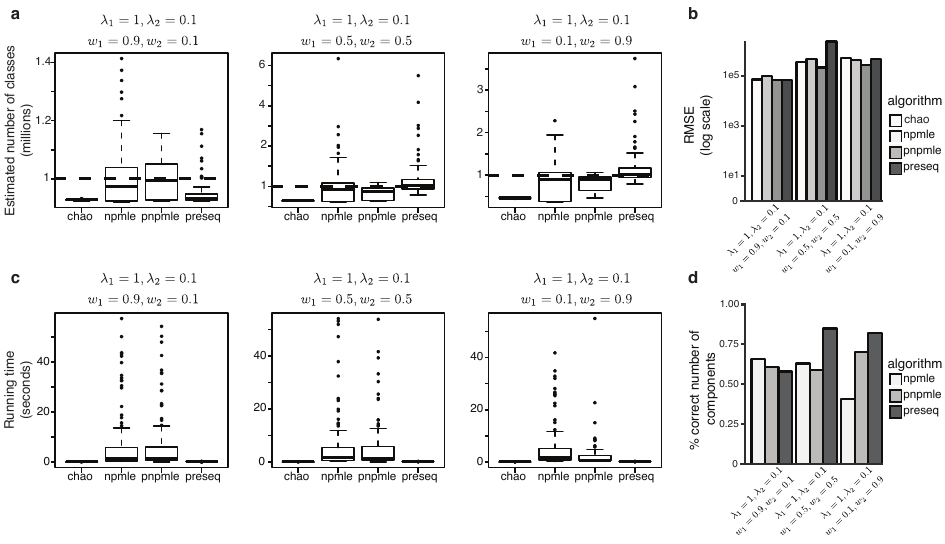}}
\caption{Comparison of chao, npmle, pnpmle, and preseq estimators for the
total number of classes when the underlying population follows a
2-component discrete distribution.  There are three cases:
$\lambda_{1} = 1, \lambda_{2} = 0.1$ and $\pi_{1} = 0.9, \pi_{2} = 0.1$;
$\lambda_{1} = 1, \lambda_{2} = 0.1$ and $\pi_{1} = 0.5, \pi_{2} = 0.5$;
and $\lambda_{1} = 1, \lambda_{2} = 0.1$ and $\pi_{1} = 0.1, \pi_{2} = 0.9$.
\textbf{a} Boxplots of the estimated number of classes.
\textbf{b} Root mean square error (RMSE) of the estimators.
\textbf{c} Boxplot of the running times.
\textbf{d} The percentage of times the npmle, the pnpmle,
and preseq chose the correct number of components in
the underlying mixture.}
\label{fig:2comp}
\end{figure}

For the two component mixture we considered three basic
cases: the more abundant classes make up a
majority of the population ($\lambda_{1} = 1, \lambda_{2} = 0.1$
and $\pi_{1} = 0.9, \pi_{2} = 0.1$);
the population is nearly equally split between a low abundance
portion and a high abundance portion
($\lambda_{1} = 1, \lambda_{2} = 0.1$ and $\pi_{1} = 0.5, \pi_{2} = 0.5$);
and the least abundant classes make up a majority of the population
($\lambda_{1} = 1, \lambda_{2}  = 0.1$ and $\pi_{1} = 0.1, \pi_{2} = 0.9$).

Figure~\ref{fig:2comp} shows the comparison of the
estimators.  We note several things,
in the latter two cases the median of the preseq estimates
is very close to the true number of classes but the RMSE is very high.
This is because in a few number of cases preseq vastly overestimates
the number of unobserved classes, indicating that the
ill-conditioning of the estimator is a problem.
We will discuss a bootstrapping strategy to deal with this
issue in the next section.
In the first case, the median of the pnpmle estimates is closest
to the true number of classes but preseq has the lowest
RMSE.
To achieve a low RMSE, it may be a viable strategy to consistently underestimate
the number of classes to decrease the variance of the estimator.
We see this in the first case studied.  The preseq estimator choose 2
components in 63 out of the 100 simulations, and defaulted
to the chao estimator in the other 37 simulation.

One very noticeable difference between the estimators is the running
time.
The chao estimator will of course be very fast to compute, but there is a
dramatic difference in running time between the moment-based
estimators and the maximum likelihood-based estimators.
The large running time of the maximum likelihood-based methods
will make it difficult to construct efficient bootstrap-aggregated
estimators.
On the other hand, it will be quite easy to construct bootstrap-aggregated
estimators for the moment-based estimators.

\subsection{Three component mixture abundance distributions}
\label{sec:3comp}

\begin{figure}[t!]
\centering{\includegraphics{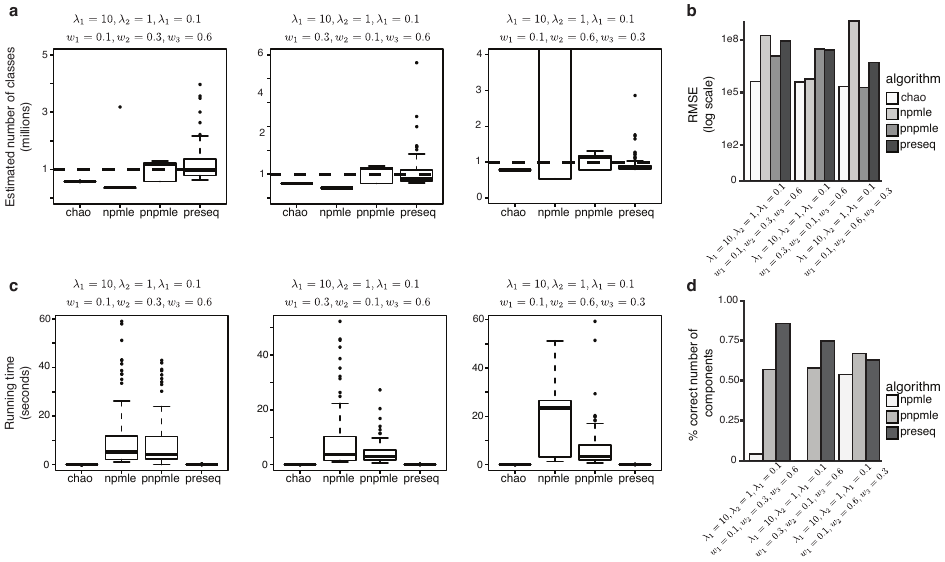}}
\caption{Comparison of chao, npmle, pnpmle, and preseq estimators for the
total number of classes when the underlying population follows a
3-component discrete distribution.  We considered three cases:
$\lambda_{1} = 10, \lambda_{2} = 1, \lambda_{3} = 0.1$ and $\pi_{1} = 0.1, \pi_{2} = 0.3, \pi_{3} = 0.6$;
$\lambda_{1} = 10, \lambda_{2} = 1, \lambda_{3} = 0.1$ and $\pi_{1} = 0.3, \pi_{2} = 0.1, \pi_{3} = 0.6$;
and $\lambda_{1} = 10, \lambda_{2} = 1, \lambda_{3} = 0.1$ and $\pi_{1} = 0.1, \pi_{2} = 0.6, \pi_{3} = 0.3$.
\textbf{a} Boxplots of the estimated number of classes.
\textbf{b} Root mean square error (RMSE) of the estimators.
\textbf{c} Boxplot of the running times.
\textbf{d} The percentage of times the npmle, the pnpmle,
and preseq chose the correct number of components in
the underlying mixture.}
\label{fig:3comp}
\end{figure}

With a three component mixture, there are many cases we could
possibly consider.
We focus on the following three cases:
$\lambda_{1} = 10, \lambda_{2} = 1, \lambda_{3} = 0.1$ and $\pi_{1} = 0.1, \pi_{2} = 0.3, \pi_{3} = 0.6$;
$\lambda_{1} = 10, \lambda_{2} = 1, \lambda_{3} = 0.1$ and $\pi_{1} = 0.3, \pi_{2} = 0.1, \pi_{3} = 0.6$;
and $\lambda_{1} = 10, \lambda_{2} = 1, \lambda_{3} = 0.1$ and $\pi_{1} = 0.1, \pi_{2} = 0.6, \pi_{3} = 0.3$.

Figure~\ref{fig:3comp} shows the comparison of the
estimators in the three 3-component cases
considered.
The median of the preseq estimates
is the closest to the true number of classes.
But what figure~\ref{fig:3comp}\textbf{a} is missing is that largest
estimate from preseq is on the order of the hundreds of millions
for all three cases.
This problem is not unique to moment-based estimation, as the maximum of
the pnpmle estimates is on the order of hundreds of millions in the first
two cases and the maximum of the npmle estimates is on the order of billions in
the first and last cases.  In fact, in the last case the median of the npmle estimates
is approximately 1.3 billion.
This is why in all cases the chao estimator vastly outperforms all other methods
in terms of RMSE.

To alleviate the issue of infrequently large estimates from an otherwise
well performing estimator, which seems to be a common problem in species estimation,
a bootstrap-aggregating (bagging) approach
was proposed by \cite{kuhnert2008bagging}.
We will discuss such an approach for the
moment-based estimator in the next section.

\section{Bagging for more accurate estimates of the missing number of classes}
\label{sec:bagging}

Consider a full set of count frequencies
$n_{0}, n_{1}, n_{2}, \ldots$. The likelihood of this set is equal
to
\[
\binom{S}{n_{0}, n_{1}, \ldots} \prod_{j = 0}^{\infty} (f_{j})^{n_{j}},
\]
with $f_{j} = \int_{0}^{\infty} e^{-\lambda} \lambda^{j} / j ! d \mu (\lambda)$.
As usual~\citep{sanathanan1972estimating}, we split the
likelihood into two parts: the unobserved portion and the observed portion
as follows,
\begin{align}
\binom{S}{n_{0}, n_{1}, \ldots} \prod_{j = 0}^{\infty} (f_{j})^{n_{j}} &=
\binom{S}{n_{0}} f_{0}^{n_{0}} (1  - f_{0})^{S - n_{0}}
\, \cdot \, \binom{D}{n_{1}, n_{2}, \ldots}
\prod_{j = 1}^{\infty} \bigg( \frac{f_{j}}{1 - f_{0}} \bigg)^{n_{j}}
\notag \\
&= L (n_{0}) \cdot L (n_{1}, n_{2}, \ldots).
\label{like_docomp}
\end{align}

Note that the observed data only enters the
likelihood above through the latter part
$L(n_{1}, n_{2}, \ldots)$.
This indicates that we can take bootstrapped samples
as multinomial samples with size parameter $D$ and
probabilities $n_{1} / D, n_{2} / D, \ldots$,
rather than sampling and the individual
level.
This will speed up the bootstrapping tremendously, as the
number of items to sample per bootstrap
is $D$, rather than $N$, and it's always the
case that $D \leq N$.

To alleviate the problem of extreme estimates,
we take the median of the bootstrap estimates
as the bagged estimator.





\section{Applications}
\label{sec:app}

It is difficult and expensive to construct real experiments
where the underlying population is known.
One example for capture-recapture experiments
is \cite{carothers1973capture}, where the author used a known
population of taxicabs and simulated capture by recording
observed registration numbers during a capture period.
For the large populations we wish to study, such experiments
would be exorbitantly expensive.


We instead take a simulation based approach, using theoretical models
proposed for our applications.
We examine two cases where the total number of
species is of great interest:
\begin{itemize}
\item T-Cell Receptor (TCR) repertoire, which are subject to depletion
due to infection and age, and we can estimate the effect age
has on reducing TCR repertoire;
\item the total number of genes present in a cell of 
a single cell RNA sequencing (scRNA-seq) experiment, 
as this indicates cell quality and can represent a
batch effect that will bias analysis~\citep{hicks2017missing}.
\end{itemize}

For comparison,
we applied our moment-based estimator along with
several other non-parametric species richness estimators.
Specifically we applied Chao's
original lower bound~\citep{chao1984nonparametric},
the duplicate fraction-based estimator of
\cite{chao2002estimating},
the jackknife estimator~\citep{burnham1978estimation},
the abundance coverage-based estimator (ACE)
\citep{chao1992estimating},
the penalized non-parametric maximum likelihood
estimator~\citep{wang2005penalized},
and the ratio regression-based method of
\cite{willis2015estimating}.
For the first five estimators we used the
R package SPECIES~\citep{wang2011species}.
For the ratio regression estimator we used the
R package breakaway.

\subsection{TCR repertoire}

\begin{figure}[t!]
\centering{\includegraphics{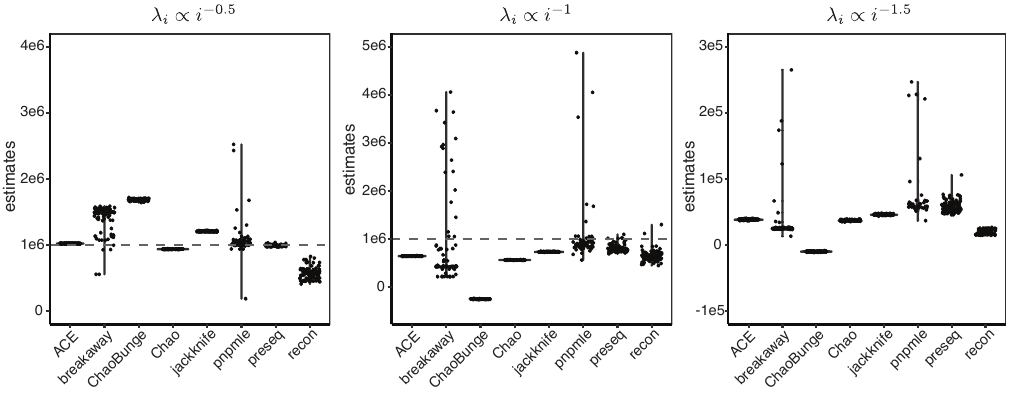}}
\caption{Violin plots of the estimated total size of the
TCR repertoire based on 100 independent samples
of approximately one million sampled T-Cell receptors for
power law distributed abundances with parameter
$\alpha = 0.5, 1, 1.5$.  Dashed line indicates the true
size of the TCR repertoire.
Some estimates are missing because they are out of the
figure boundaries.
See Supplementary Figure~\ref{power_law_figure_log_scale}
for the estimates on the log scale.}
\label{power_law_figure}
\end{figure}

Interrogations of TCR repertoire typically reveal a long
tails of highly abundant and, at the same time,
a large number of rare
receptors, e.g. Table~\ref{tcr_freq}.
This behavior is typical of power law distributions,
and recent work has shown that competitive effect for
TCR binding may explain this behavior
\citep{desponds2016fluctuating}.

We took a theoretical population of one million TCRs, with
abundances $\lambda_{i} \propto i^{- \alpha}$ for $i = 1, \ldots, 10^6$
and $\alpha = 0.5, 1, 1.5$.
We set the average number of total TCRs sampled equal to
one million, i.e. $\sum_{i} \lambda_{i} = 10^{6}$.
We then sampled 100 independent samples and compared the
estimates obtained from
ACE, breakaway, ChaoBunge, jackknife, pnpmle, preseq,
and recon, a new non-parametric estimator designed
specifically for estimating TCR repertoire~\citep{kaplinsky2016robust}.

The results of these simulations are shown in Figure~\ref{power_law_figure}.
In all three cases considered, preseq estimates were the closest
to the true number of species on average,
as well as the lowest RMSE of any estimator.
This indicates that increased variance from using more moments is
offset by increased accuracy, particularly compared to Chao's
lower bound.
In the cases $\alpha = 0.5$ and $\alpha = 1$, preseq estimates are
not a strict lower bound, as Chao's estimates are, but we believe
that the increased accuracy is beneficial enough to compensate for
this drawback.

Some of the estimators, such as breakaway or pnpmle, had extremely
large variation in their estimates, with estimates ranging over several
orders of magnitude (Figure~\ref{power_law_figure_log_scale}).
This may be an indication that the models assumed by these estimators
are not able to account for populations with long tailed abundances,
typical of power-law distributions~\citep{newman2005power}.
Other estimators, such as ChaoBunge, had negative estimates,
as others have previously found~\citep{wang2005penalized}.
This may indicate that the models these estimators use are not
flexible enough to handle power law-like populations,
despite their previous use in estimating TCR repertoire~\citep{qi2014diversity}.

\begin{figure}[t!]
\centering{\includegraphics{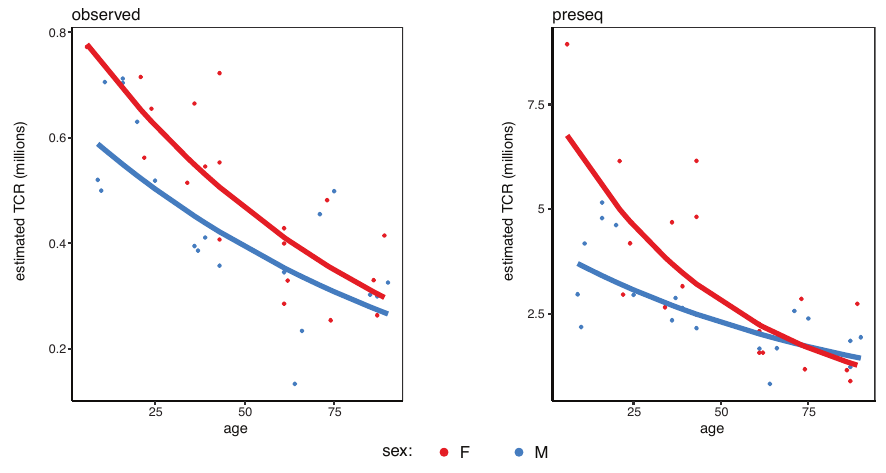}}
\caption{Estimated T-Cell repertoire as a function of age for the
 observed TCR and preseq estimates, along with estimated
 log-linear regressions for males and females subject separately.}
\label{TCRvsAge}
\end{figure} 

Now we apply our method to real data from \cite{britanova2014age},
looking at the decrease in T-Cell repertoire as a function of age.
We assume that total TCR decreases exponentially with age,
as a linear regression may result in negative estimates of the
total repertoire.
We compare female and male TCR as a function of age, looking
at the original observations and the preseq missing species-corrected
estimates, shown in Figure~\ref{TCRvsAge}.
Other estimators are show in Supplementary Figure~\ref{TCRvsAgeSupp}.
We see that after correction for the missing repertoire,
female TCR is predicted to be lower on average than male TCR
at around 75 years old.
Without correction for the missing repertoire, the
repertoire for females is not predicted to be lower
than the male repertoire at any age considered.
The consequence of lower T-Cell repertoire at advanced
age is increased risk of infection~\citep{yager2008age,blackman2011narrowing},
and such insights may guide decisions of treatment
and outreach to prevent infections.
We should note that the above is an example
of exploratory analysis and
a true test for the difference needs to take
into account both the observed variance and
variance from the estimated missing species
\citep{willis2017improved}.

\subsection{Single cell RNA sequencing}

Single-cell RNA sequencing (scRNAseq) is a new technology to
interrogate the transcriptome of individual cells.
Unfortunately, most assays can not capture the full transcriptome in an unbiased manner.
Low abundant genes are more difficult to capture and are said to 
suffer from dropout when they are present in the original
cell but are not captured by the assay~\citep{hicks2017missing}.
These experiments also exhibit huge deviations from uniformity
due to technical biases.
It is therefore difficult to determine whether a gene is missing
due to dropout or due to insufficient sequencing depth.

Estimating the dropout rate is critical to the normalization
of scRNAseq data~\citep{pierson2015zifa,lun2016pooling}
and is equivalent to estimating
the number of genes that are present in the experiment 
but were not sampled due to insufficient sequencing depth.
We can apply species sampling models to estimate
the dropout rate.  
Our previous experience has shown the value in non-parametric
models for sequencing experiments~\citep{daley2013predicting},
so we will test our proposed estimator and other non-parametric
estimators
to the problem of estimating the dropout rate in simulated
scRNAseq experiments.

We use the simulation framework of \cite{lun2016pooling}, with some
small modifications proposed by~\cite{hicks2017missing} to introduce
batch-related dropout effects.
We assume that we measured 20,000 genes in 1,000 captured cells.
We assume that the population consists of two subpopulations,
each making up half of the cells.
These may represent different cell types or another biological 
condition.  
We assume that the count of gene $i$ in cell $j$ obtained from the scRNAseq
experiment is a log-Normal-Poisson random variable
with mean equal to $ d \theta_{j} \lambda_{ij}$ times 
independent log normal random variables
with parameters $\mu = -0.5$ and $\sigma = 1$.
We assume that $\theta_{j}$'s are independent log-normal
random variables with parameters $\mu = -2.5^2/2$ and $\sigma = 2.5$.
We assume that $\lambda_{ij} = \lambda_{i0} \phi_{ij}$,
$\lambda_{i0} \sim \text{Gamma}(0.1, 0.1)$, $d = 0.05$,
and $\phi_{ij}$ is a sub-population specific value.
We assume that for cells in the first sub-population $\phi_{ij} = 1$ for
all genes and that for cells in the second population 
$\phi_{ij} = 5$ for $10 \%$ of the genes,
$\phi_{ij} = 0.2$ for another $10 \%$ of the genes, 
and $\phi_{ij} = 1$ for the remainder.
The former two sets of genes represent condition specific 
differentially expressed genes.
These parameters were chosen to simulate the properties
of droplet-based scRNAseq experiments,
as these allow for more cells to be captured but
tend to be sparser than microfluidic-based experiments.

Following the dropout model of \cite{hicks2017missing},
we assumed that counts dropout
with a probability that is logistic function of the baseline,
with batch specific baseline values.
Specifically we assume that the baseline probability
$1 / (1 + \exp (-\beta_{0ij}) )$ follows a Beta$(2, 8)$ for half the population
and a Beta$(2, 38)$ for the other half, independent of condition.
The dropout probability is then equal to 
\[
p_{ij} = 1/ \big(1 + \exp \big(- \beta_{0ij} - 0.5 (\theta_{j} \lambda_{ij} - \bar{\theta_{\cdot} \lambda_{\cdot \cdot}}) \big) \big).
\]
This simulates the condition where the population was processed in 
two batches, each with a batch specific dropout rate.  
After standard filtering (at least 50 counts per cell, 10 counts
per gene, and genes must appear in at least 3 cells), we obtained
a counts matrix of 436 cells and 5,262 genes
(full details can be found at \url{https://github.com/timydaley/SingleCellWeights}).

\begin{figure}[t!]
\centering{\includegraphics{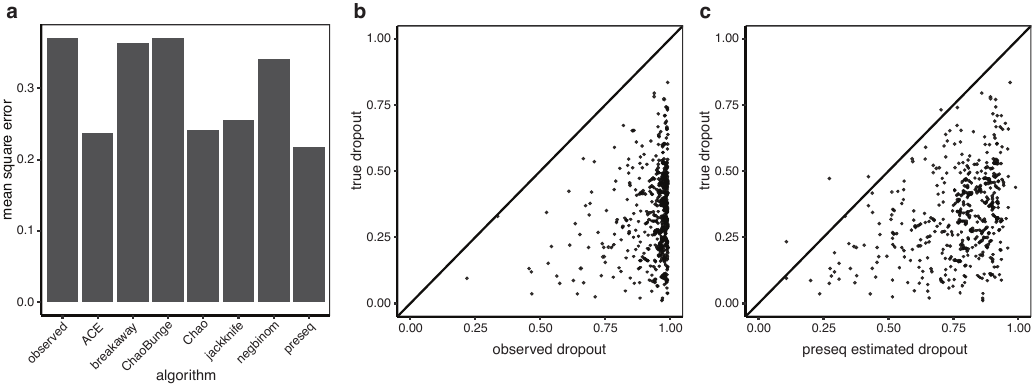}}
\caption{Estimating the number of unobserved genes improves
dropout estimation.
\textbf{a} Correlation of the corresponding estimated dropout rates 
to the true dropout rates.
\textbf{b} Scatter plot of the observed dropout rate versus the true dropout rates.
The dashed line indicates perfect estimation of the dropout rate.
\textbf{c} Scatter plot of the preseq estimated dropout rates versus the true dropout
rates.  
The other estimators are shown in Supplementary 
Fig~\ref{scRNAseqLogNormalModelSupplementary}.
The improvement in estimation comes mostly from cells with high observed
dropout.
}
\label{scRNAseqLogNormalModel}
\end{figure}

The log-Normal Poisson is common for RNA-seq data,
{\em e.g.} \cite{jia2017accounting}, but initial tests indicate that 
the dropout and low sequencing depth of scRNA-seq experiments
result in extremely high estimates of dropout (more than the number of genes).
Accordingly, many groups have developed tools that extend
models to account for dropout, {\em e.g.} 
\cite{kharchenko2014bayesian} and \cite{pierson2015zifa}.
Both of these do not allow for cell specific dropout rates, 
but recently researchers have recognized the need for 
estimating the dropout rate at the cell level~\citep{risso2018general}.
We investigated how non-parametric species estimation can
improve the estimation of the dropout rate.
We will use the same methods
as in the previous section, excluding the pnmle and recon,
as these methods are suitable for application to a single sample
but take too long to apply to the number of number of samples (cells)
obtained from a single cell RNA-seq experiment.
We also added comparisons to a Negative Binomial parametric
model, where the positive counts are assumed to arise from
a zero-truncated Negative Binomial distribution.
This represents a case where a close but incorrect parametric
model is used.  
We used the EM algorithm available in the preseqR package
\citep{dengpreseqr} to fit the parameters of the Negative Binomial model.
For methods that estimated a dropout rate less than zero, 
as occurs when the number of estimated genes is greater
than the known number of genes,
we set the dropout rate equal to the observed dropout rate.

Nearly all of the methods improve the estimated
dropout rate (Fig~\ref{scRNAseqLogNormalModel}\textbf{a}).
The mean square error improved in all cases.  
The best performing methods in terms of lowest mean
square error were the Chao lower bound, the ACE, 
and preseq (Fig~\ref{scRNAseqLogNormalModel}\textbf{a}).  
All three improved the estimated dropout rate 
considerably better than all other tested methods,
with preseq showing the highest performance.
Most of the improvement was in the cells with 
very high observed dropout rates 
(Fig~\ref{scRNAseqLogNormalModel}\textbf{b} and \textbf{c}).
These cells present the most problems to investigators,
as it is extremely unlikely that the observed sparse gene
expression is due to biological factors.

\begin{table}[t!]
\begin{center}
\begin{tabular}{r r r r r}
\hline
method & $p_{\text{adj}} < 0.01$ & $p_{\text{adj}} < 0.05$ & $p_{\text{adj}} < 0.1$ \\
\hline
un-corrected & 402 & 585 & 768 \\
array weights & 221 & 275 & 314 \\
observation weights & 162 & 256 & 345 \\
preseq dropout & $\boldsymbol{643}$ & $\boldsymbol{771}$ & $\boldsymbol{771}$ \\
\hline 
\end{tabular}
\caption{The number of genes correctly identified as differentially expressed with the correct sign at Bejamani-Hochberg adjusted $p = 0.01, 0.05, $ and $0.1$ for un-corrected, array weight-corrected, observation weights-corrected, and preseq dropout-corrected differential expression.}
\label{diff_expr_table}
\end{center}
\end{table}

To illustrate the application of estimating the dropout
to single cell RNA-seq, we use one minus the
preseq-corrected dropout rate as cell weights in
differential expression and show that this improves
the performance of standard RNA-seq differential expression
tools.
The idea is that we want to up-weight cells that we trust more
and down-weight cells that we trust less.
Intuitively, we should trust cells with higher dropout
less than we trust cells with a low dropout rate.
If the differential expression is independent or orthogonal
of the dropout rate, then this should improve
the differential expression analysis.
This is similar to the idea of using observation weights
for scRNA-seq analysis presented in \cite{van2018observation}
but more akin to the approach presented by \cite{liu2015weight},
since the former weights every count individually and we
are weighting the cells (i.e. samples).
We applied this to the same data set as above,
using edgeR/limma~\citep{robinson2010edger,ritchie2015limma}
to compute differential expression.
We used the preseq corrected dropout weights as 
input for the array weights in the standard pipeline.
We compared the preseq corrected
differential expression results to the
uncorrected differential expression results, 
the array weights corrected differential expression
results using the limma computed array weights,
and differential expression results using 
observation weights from ZINB-WaVE
\citep{risso2018general}.  
At B-H adjusted p-values of $0.01$, $0.05$, and $0.1$,
the preseq corrected results identified the
highest number of truly differentially expressed
genes (Table~\ref{diff_expr_table},
without a corresponding increase in the false
positive rate
(Fig~\ref{diff_expr_tpr_fpr}).

These results indicate that the use of species
estimation for cell-specific dropout correction
can improve the analysis for differential
expression analysis for single cell
RNA-seq.  
Though a caveat must be made.  
If the conditions significantly differ in their
dropout rates, possibly due to batch effects,
then dropout correction may bias the differential
expression analysis. 
This is something that the researcher should check 
early in the analysis, and it may indicate deeper
problems with the data than can be corrected

\section{Discussion}
\label{sec:discuss}

We presented a moment-based method
for estimating the number of unobserved
species in a large and heterogeneous population.
This builds on the work of \cite{harris1959determining}
and extends the estimator of \cite{chao1984nonparametric}
to modern large-scale sampling experiments,
such as those that arise from high-throughput DNA
sequencing.
These experiments explore populations
that are sizable and extremely diverse.
To show highlight the performance of our estimator,
we compared our method to other
non-parametric estimators in two
specific applications of species richness estimation
in modern biology: T-Cell receptor repertoire
and single-cell RNA-seq.
We showed that in these two applications
the use of our method can improve downstream
inference.

We note one caveat of our method.
Following the framework of \cite{harris1959determining},
our method can be used to obtain bounds on a broad range of
functionals of the abundance distribution.
However, we believe that it should not
be used to recover the abundance distribution.
Methods such as the npmle~\citep{norris1998non} or the
pnpmle~\citep{wang2005penalized} are specifically designed
for this task and should be used instead.
Indeed, we think that this problem is much more difficult
and one benefit of our method is the avoidance of this problem.

As we mentioned briefly in section~\ref{sec:mom_spaces},
upper bounds for the number of species is infinite due
to the inclusion of the boundary point 0 in the abundance
distribution.
If the abundance distribution is assumed to have a non-zero
lower bound, then upper bounds can theoretically be obtained
via Gaussian quadrature.
Preliminary results indicate that the upper bounds are
too large to be practically useful, hence our focus on
lower bounds.

Our method is available in the open source package
preseq (\url{https://github.com/smithlabcode/preseq})
under the bound\_pop module.

\paragraph{Acknowledgements}
The authors would like to thank Kristian Lum,
Peter Calabrese, Manuel Lladser, Susan Holmes,
James Johndrow, and especially Chao Deng
for their helpful comments and discussion.
Without their help this paper would be have
been forever stuck in limbo.

\pagebreak

\bibliographystyle{natbib}
\bibliography{biblio.tex}

\newpage

\begin{center}
\textbf{\large Supplemental Figures}
\end{center}

\setcounter{equation}{0}
\setcounter{figure}{0}
\setcounter{table}{0}
\setcounter{page}{1}
\makeatletter
\renewcommand{\theequation}{S\arabic{equation}}
\renewcommand{\thefigure}{S\arabic{figure}}
\renewcommand{\bibnumfmt}[1]{[S#1]}

\begin{figure}[h]
\centering{\includegraphics[width=15cm]{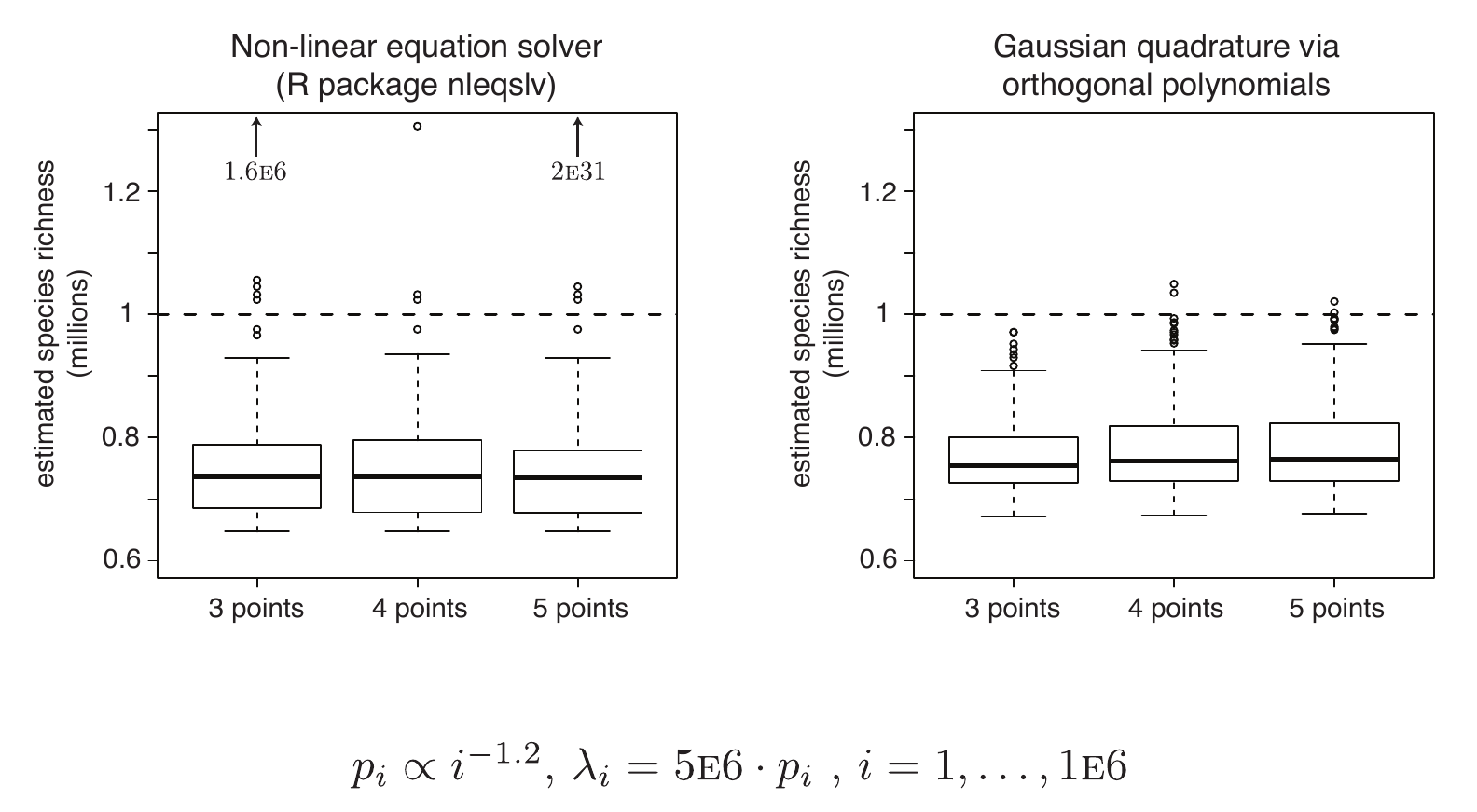}}
\caption{A comparison of directly solving the non-linear equations
\eqref{system_eqns} with a non-linear equation solver~\cite{hasselman2018package}
and via orthogonal polynomials.  We simulated 100 samples from a power-law
population and applied both algorithms using the same counts.}
\label{QuadratureOrthPolyVsNleqslv}
\end{figure}

\newpage

\begin{figure}[h]
\centering{\includegraphics{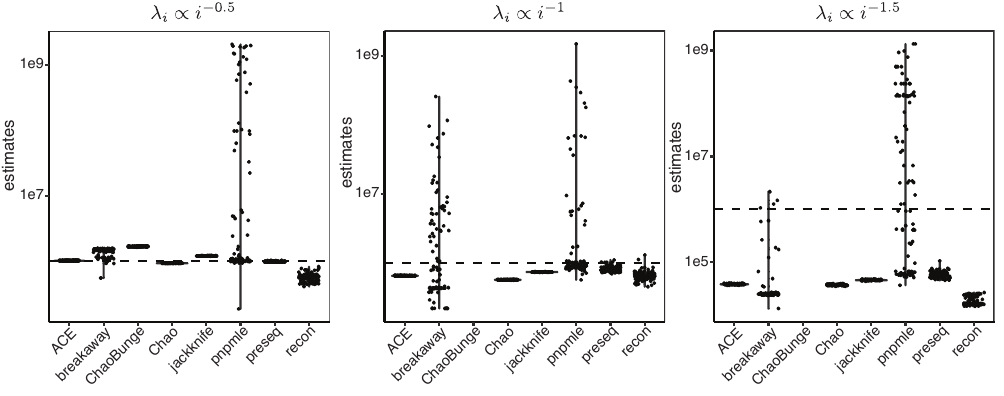}}
\caption{Violin plots of the estimated total size on the log scale of the
TCR repertoire based on 100 independent samples
of approximately one million sampled TCRs for
power law distributed abundances with parameter
$\alpha = 0.5, 1, 1.5$.  Dashed line indicates the true
size of the TCR repertoire.
Not shown are negative estimates,
but these are shown in figure~\ref{power_law_figure}.}
\label{power_law_figure_log_scale}
\end{figure}

\newpage

\begin{figure}[t!]
\centering{\includegraphics{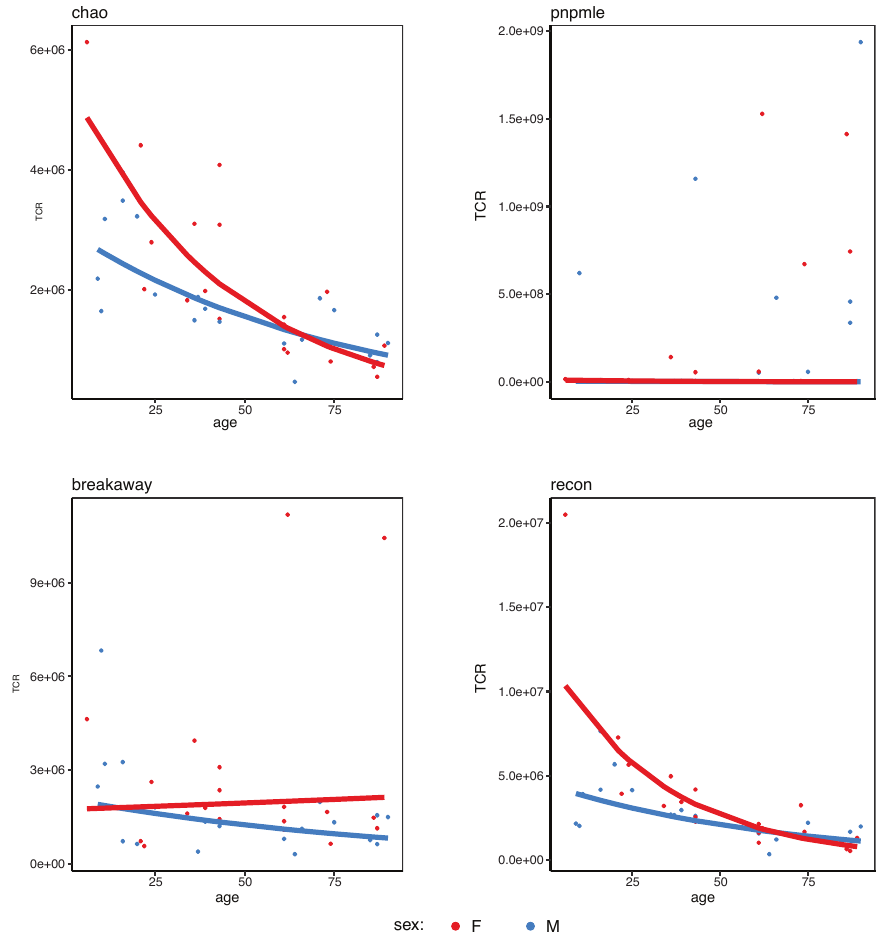}}
\caption{Estimated T-Cell repertoire as a function of age for the
 chao, pnpmle, breakaway, and recon estimates of TCR, along with estimated
 log-linear regressions for males and females subject separately.}
\label{TCRvsAgeSupp}
\end{figure}

\newpage

\begin{figure}[h]
\centering{\includegraphics{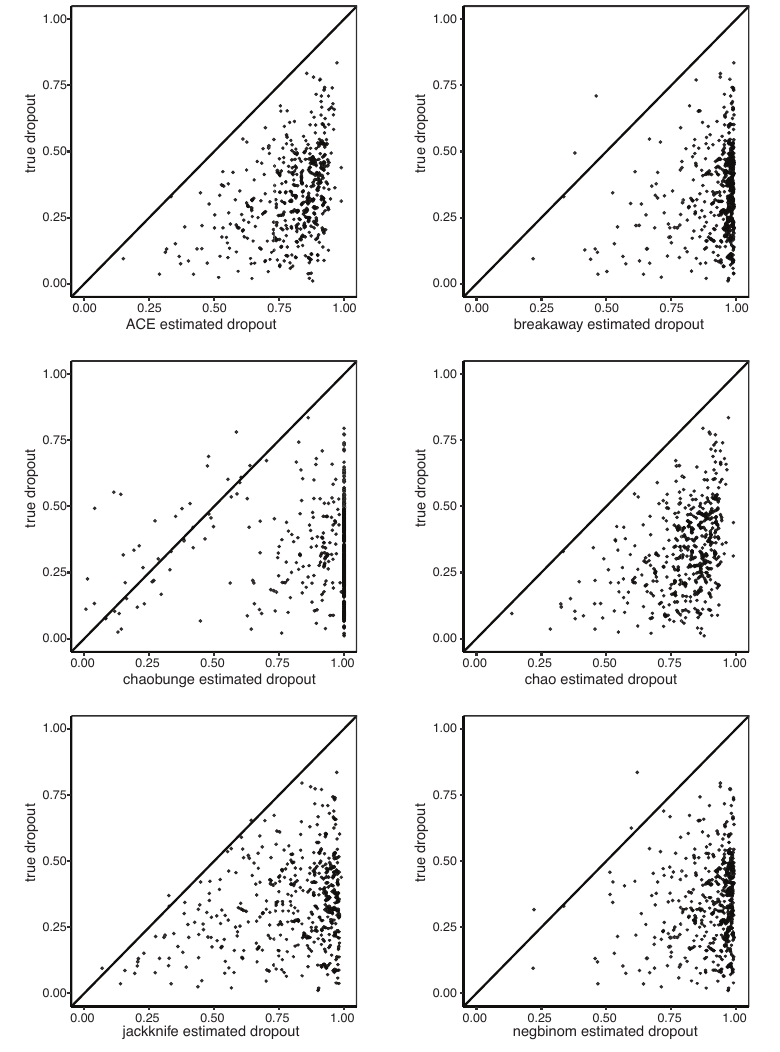}}
\caption{Plots of the estimated dropout versus the true dropout for the
Chao, ACE, ChaoBunge, breakaway, and jackknife estimators, as described in
Fig~\ref{scRNAseqLogNormalModel}\textbf{b} and \textbf{c}.
}
\label{scRNAseqLogNormalModelSupplementary}
\end{figure}

\newpage

\begin{figure}[h]
\centering{\includegraphics{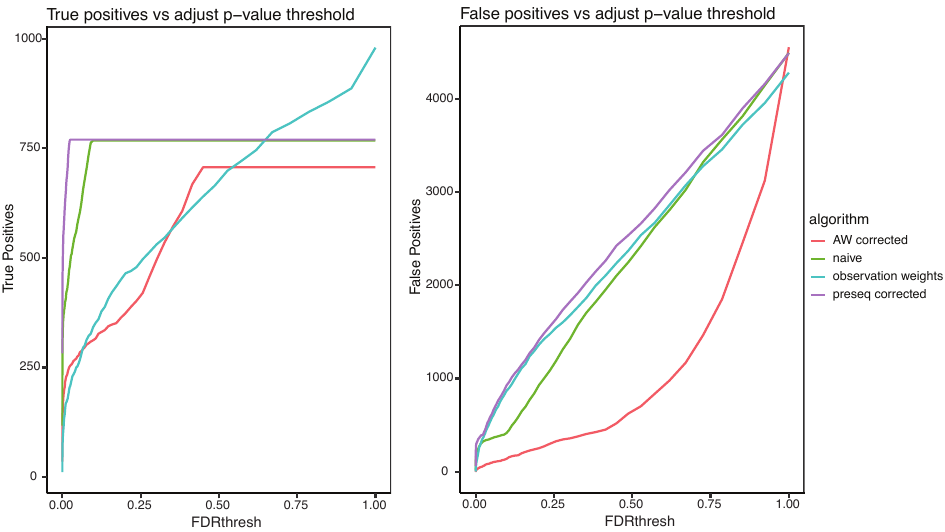}}
\caption{The number of true differentially expressed genes identified with the correct sign (left) out of a total of 1,167 truly differentially expressed genes and the number of incorrectly identified differentially expressed genes, either truly null or with the wrong sign,
as a function of Benjamani-Hochberg corrected false discovery rate.}
\label{diff_expr_tpr_fpr}
\end{figure}

\end{document}